The Enhanced Ferromagnetism of Single-Layer CrX$_3$ (X=Br and I) by Van der Waals Engineering

Hongxing Li[1,2], Yuan-Kai Xu[1,2], Kang Lai[1,2], and Wei-Bing Zhang[1,2,*]

1) *School of Physics and Electronic Sciences, Changsha University of Science and Technology, Changsha 410114, People's Republic of China*
2) *Hunan Provincial Key Laboratory of Flexible Electronic Materials Genome Engineering, Changsha University of Science and Technology, Changsha 410114, People's Republic of China*

*Correspondence to: zhangwb@csust.edu.cn

**Abstract**
The recent experimental discovery of intrinsic ferromagnetism in single-layer CrI$_3$ opens a new avenue to low-dimensional spintronics. However, the low Curie temperature *Tc* ∼ 45 K is still a large obstacle to its realistic device application. In this work, we demonstrate that the *Tc* and magnetic moment of CrX$_3$ (X=Br, I) can be enhanced simultaneously by coupling them to buckled two-dimensional Mene (M=Si, Ge) to form magnetic van der Waals (vdW) heterostructures. Our first-principles calculations reveal that n-doping of CrX$_3$, induced by a significant spin-dependent interlayer charge-transfer from Mene, is responsible for its drastic enhancement of *Tc* and magnetic moment. Furthermore, the diversified electronic properties including halfmetallicity and semi-conductivity with configuration-dependent energy gap are also predicted in this novel vdW heterostructure, implying their broad potential applications in spintronics. Our study suggests that the vdW engineering may be an efficient way to tune the magnetic properties of 2D magnets, and the Mene/CrX$_3$ magnetic vdW heterostructures are wonderful candidates in spintronics and nanoelectronics device

**Introduction**
During the past decades, a large family of 2D materials with outstanding properties have emerged as the excellent candidates in many fields, such as nanoelectronics, catalysis and energy storage.[1-4] Benefit from the atomic flat surface and the absence of dangling bond, it is easy to stack 2D materials vertically to form van der Waals (vdW) heterostructures with novel properties distinct from monolayer components.[5-8] By choosing the components and stacking styles, heterostructures with desired functions can be designed rationally. However, due to the lack of intrinsic magnetism in 2D limit, the application of 2D materials and their vdW heterostructures in spintronics has been hindered.[9-13] To overcome this problem ,extensive experimental and theoretical efforts have been devoted, such as introducing external magnetism into 2D sheet,[9,10] and seeking intrinsic 2D magnetic material.[11-13] Nevertheless, breakthrough was only achieved recently. Single-layer CrI$_3$ has been identified as a intrinsic Ising ferromagnet with out-of-plane spin direction,[14] which opens a new avenue to the development of low dimensional spintronics devices. CrI$_3$ is receiving increasing attention for its Ising type ferromagnetism and semi-conductivity, which is also suggested to be used as high-efficient magnetic tunneling barrier. The low dimensional magnetic tunneling junctions composed by few layers CrI$_3$ and graphene has been fabricated successfully by several groups,[15-18] which demonstrate high tunneling magnetoresistance even up to 1 000 000%. In addition, CrI$_3$ also provide new opportunity for the study of new type quasi-particle,[19] and the interaction between magnetism and light.[20] However, the low Curie temperature $T_c$ of 45 K14 of $CrI_3$ is still a large obstacle to its realistic device application.

Tuning the property of 2D materials is one of the most important topics in condensed physics and material science. Conventional strategies such as strain and adsorption have been applied widely to manipulate various physical properties of 2D materials including magnetism.[21,22] On the other hand, the proximity effect have also been proved to be an effective way to regulate the properties of 2D materials. The recently rising vdW heterostructures provide an ideal platform to realize the proximity effect. Since the interlayer interaction is predominated by weak vdW interaction, the common and uncontrollable interfacial defect and trap states in conventional heterostructures can be avoided. For example, the spin-orbit coupling of graphene can be enhanced by depositing them on $WSe_2$.[23] A recent experiment indicated that a ferromagnetic EuS film in contact with a topological insulator $Bi_2Se_3$ might show a largely enhanced Curie temperature and perpendicular magnetic anisotropy.[24]

In the present work, we construct a series of magnetic vdW heterostructures composed with the newly discovered 2D magnet $CrX_3$(X=Br, I) and buckled two-dimensional Mene (M=Si, Ge). Silicene and germanene are wonderful post-IV Dirac materials for their stronger spin-orbit coupling (SOC) than graphene.[25] Although the Mene based vdW heterostructures have also been investigated extensively,[26-34] the magnetic vdW heterostructures that composed by Mene and 2D magnets is still scarce. Our results predicted an enhanced ferromagnetism in $CrX_3$ by vdW Engineering. In addition, these heterostructures are also found to possess versatile electronic structure.

**Calculation method**

All spin-polarized density functional calculations are carried out using the Vienna *ab* initio simulation package (VASP)[35,36] within the projected augmented wave (PAW) method[37] and the generalized gradient approximation (GGA) in the Perdew-Burke-Ernzerhof (PBE) implementation.[38] DFT-D2 method proposed by Grimme[39] was adopt to account for the interlayer van der Waals interactions. A vacuum thickness of 15 Å is used to avoid the artificial interaction between images of heterostructures. The cutoff energy of plane wave basis is set to be 500 eV. The atomic position is relaxed until the residual forces on each atom less than $1\times10^{-2}$ eV/Å. Brillouin zone sampling is done by a $9\times9\times1$ Γ-center mesh.[40]

**Results and discussions**

The optimized in-plane parameters for $CrBr_3$, $CrI_3$, silicene and germanene are 6.345, 6.875, 3.845 and 4.017 Å, which are consistent with previous reports.[11,29,41] The $Ge/CrI_3$, $Si/CrI_3$ and $Si/CrBr_3$ heterostructure slabs are constructed by $\sqrt{3}\times\sqrt{3}$ Mene supercell to match the $CrX_3$ single-layer, and the lattice mismatch is 1.2%, 3.2% and 5.0%, respectively.

According to the stacking style between Mene and $CrX_3$, three high symmetrical positions namely TX, TbX and TuX, are considered in our calculations. Hexagonal rings with different colours are also shown in Figure 1(a) for a clear illustration. For position TX, all M atoms are right above the top of all X atoms. For position TbX/TuX, half M atoms locate above the bottom/upper X atoms. TX can be converted to TbX (TuX) by sliding about 1/3 $CrX_3$ lattice parameter along zigzag (armchair) direction. Since Mene is bulked, two different configurations can constructed at each position. For instance, at TX, we can place the up or down M atom right above upper X atom to form the two configurations labeled as TX_up and TX_dn. Therefore, there are total six different stacking configurations as shown in Figure 1.

The equilibrium interlayer distance $d$, and binding energy $E_b$ per M atom of six stacking configuration are listed in Table I and Figure 2. $E_b$ is calculated by $E_b=(E_M+E_{CrX3}-E_{Heter})/n$, where $E_M$, $E_{CrX3}$, and $E_{Heter}$ are the total energy of Mene, CrX$_3$ and heterostructure, respectively. As we can see, TbX up is the most energetically favorable configuration for all heterostructures with the largest $E_b$ of 150, 95 and 66 meV for Ge/CrI$_3$, Si/CrI$_3$ and Si/CrBr$_3$, respectively. Ge/CrI$_3$ and Si/CrBr$_3$ have strongest and weakest interlayer interaction. However, the interlayer interactions of Mene/CrX$_3$ are still weaker than the system of Mene/InSe in the same PBE-D2 level.[29] At each position for all heterostructures, the change of $E_b$ and $d$ that result from vertical inversion between up and down M atom is very small, which range from 0.01 Å to 0.11 Å and from 3 meV to 9 meV, respectively.

To gain further insight into the interlayer interaction, we calculate the charge density difference (CDD). As shown in Figure 3, the distinct configurations-dependent charge redistribution of Ge/CrI$_3$ system can be found, and small structure change can generate different charge redistribution. More interestingly, the vertical inversion between up and down M atoms in heterostructure gives rise to significantly different CDD, although tiny energy change was found above. For example, there is charge accumulation in interlayer space of Ge/CrI$_3$ with TX_dn configuration, but no charge accumulation is found in case of TX_up. At TuX_up, the Ge atoms and upper I atoms exhibit significant self redistribution, and some charge accumulate at interlayer spaces. However, at TuX_dn, there is no charge redistribution at the corresponding spaces. Nevertheless, regardless of the configuration, there is always charge accumulation at Cr atom. We further calculate the charge transfer quantitatively in these heterostructures by Bader charge analysis,[42] which are listed in Table I and Figure 2. We find CrX$_3$ always get electrons from Mene, and results in n-doping. This is also in accord with the fact that the electron affinity of CrI$_3$ (CrBr$_3$), 4.68 eV (4.91 eV), is larger than the work function of germanene (silicene), 4.34 eV (4.64 eV).[43] As the charge redistributions arise in heterostructures, it is interestingly to look into the alteration of the magnetic properties of CrX$_3$. The average magnetic moments of Cr atoms in heterostructures for different stacking configurations are 3.156, 3.102 and 3.047 $\mu B$ for Ge/CrI$_3$, Si/CrI$_3$ and Si/CrBr$_3$, respectively, larger than the values of 3.098 (3.008) $\mu B$ in single-layer CrI$_3$ (CrBr$_3$). The spin-dependent CDD are calculated to uncover the reason, and Figure 3S in SI depict the result for Ge/CrI$_3$ in TuX_dn as a demonstration. We find the Cr atom gain spin-up charge, while the spin-down charge is lost. In single-layer CrX$_3$, the $d$ electrons of Cr atoms just occupy spin-up direction orbits.[11] Therefore, the accumulation of spin-up charge results in a larger magnetic moment than pristine CrX$_3$. This similar situation is also reported in the system of iron(II) phthalocyanine molecule on Bi$_2$Te$_3$ substrate, in which the molecular magnetic moment will increase as charge transfer from substrate.[44] On the other hand, Mene was magnetized due to proximity effect, and Ge atoms can acquire a largest magnetic moment of 0.035 $\mu B$ that antiparallel to Cr atom at TuX up in Ge/CrI$_3$. This value is much larger than the 0.003 $\mu B$ of Se atom in Bi$_2$Se$_3$/CrI$_3$.[45] In order to determine the magnetic ground state and evaluate magnetic stability of CrX$_3$, we calculate the exchange energy $\Delta E$ of CrX$_3$ in heterostructure, $\Delta E=E_{AFM}-E_{FM}$. The results are listed in Table 1 and Figure 3(d). The $\Delta E$ of single-layer CrX$_3$ was also calculated, and the result is 39(29) meV for CrI$_3$ (CrBr3). As we can see, $\Delta E$ varies with systems and configurations, and range from 53 to 61, 38 to 41, 47 to 53 meV for Ge/CrI$_3$, Si/CrI$_3$ and Si/CrI$_3$, respectively. The $\Delta E$ of CrX$_3$ in the heterostructures is usually larger than the freestanding single-layer one. Compared with free-standing single-layer CrX$_3$, $\Delta E$ in Ge/CrI$_3$ can increase maximally by 56% at TbX_up and TuX_up, and in Si/CrBr$_3$ it can increase maximally by 83% at TuX_dn. According to mean-field theory, the Curie temperature $T_c$ is expected to be proportional to $\Delta E$. Using the experimental $T_c$ of single-layer CrI$_3$ 45 K,[14] the highest $T_c$ of CrI$_3$ is evaluated to 70 K, close to the liquid nitrogen temperature. It is worthy to note that the $\Delta E$ of Ge/CrI$_3$ is always larger than Si/CrI$_3$, indicating the overlaying of germanene on CrI$_3$ will lead to a more stable ferromagnetism than silicene. The enhancement of ferromagnetism in heterostructures can be ascribed to charge transfer from Mene to 2D magnets as discussed above. This is consistent with previous study of Wu et al.,[46] in which they found the charge doping can steadily enhance the ferromagnetism of CrI$_3$. The similar enhancement of ferromagnetism in magnetic single-layer induced by substrate is also found in the case of VSe$_2$/MoS$_2$ by experiment recently.47

We also calculate the density of states (DOS) of these systems at different configurations. The DOS of Ge/CrI$_3$ for six configurations are presented in Figure 4 and other configuration can be found in Figure S4 and S5 in SI. Compared with the single-layer CrX$_3$, the fermi level of these heterostructures moves toward to conduction band, confirming the n-doping. The M $p$ states hybrid with Cr $d$ and X $p$ states, as they show some identical peaks. However, the top of valence band is dominated by Ge $p$ states, while the bottom of conduction band is dominated by Cr $d$ and X $p$ states. The Dirac cone of Mene was destroyed seriously, even similar to the situation on metal substrate.[26-28] Nevertheless, it is very different from Mene on other non-magnetic insulating 2D substrates, such as MoS$_2$, WSe$_2$ and MX (M=Ga, In; X=S, Se, Te).[29-32] The Dirac cone of Mene was found to be well preserved, even though the binding strength between Mene and these substrates may be stronger than on CrX$_3$. Unlike on usual nonmagnetic substrates, the DOS in spin up and down channels of Mene around Fermi level are asymmetrical and exhibiting strong spin polarizations, as a consequence of magnetic proximity effect. Therefore, in addition to hybridization and asymmetrical potential, the exchange splitting induced by ferromagnetic substrate may be another reason for the destroying of Dirac cone. The spin-polarized band structures of these heterostructures are calculated to understand the electronic properties of heterostructures. The band structures are shown in SI, and we summarize the characters of all band structures in Figure 5. As we can see, the band structures of each system at different configurations show significant difference. For example, Ge/CrI$_3$ manifest as semiconductor with a band gap of 165 meV at TX_up, while at TbX_up it become a half-metal with a spin gap of 200 meV. At TX_dn, Ge/CrI$_3$ is semiconductor with a rather small band gap of 25 meV, the 1/6 times as TX_up. However, the structural difference between TX_up and TX_dn is just the inversion of up and down Ge atoms. Apart from semiconductor and halfmetal, Ge/CrI$_3$ can be metal at TuX_dn. Si/CrI$_3$ can be metal and semiconductor, and the largest gap is 218 meV at TX_up and the smallest one is 54 meV at TbX_up. Different from Ge/CrI$_3$ and Si/CrI$_3$, Si/CrBr$_3$ is always half-metallic, but the spin gap vary with stacking configuration. At TbX_dn, Si/CrBr$_3$ has smallest spin gap of 35 meV, and the largest gap is 290 meV at TuX_dn. The strong configurations-dependent electronic properties of heterostructures are also in consistent with the fact that the CDD changes discussed above.

**Summary**

In summary, the first principle calculations are employed to study the structural, magnetic and electronic properties of Ge/CrI$_3$, Si/CrI$_3$ and Si/CrBr$_3$ magnetic vdW heterostructures. Electrons will transfer from Mene to CrX$_3$, leading to n-doping in CrX$_3$. The n-doping will enhance the ferromagnetism of 2D magnets, as exchange energy of CrI$_3$ and CrBr$_3$ increased maximally by 56% and 83%, respectively. The $T_c$ of CrI$_3$ can increase to 70 K. Our study indicates the van der Waals engineering may be an effective way to tune the magnetic and structure property of 2D magnet. Moreover, the electronic structures of these heterostructures show strong dependence on stacking configurations, as they can be metal, or semiconductor and half-metal with different band gaps. The multiple properties make these magnetic vdW heterostructures have diversified potential applications in spintronics and nanoelectronics.

**ACKNOWLEDGMENTS**
This work was supported by National Natural Science Foundation of China (Grant No.11847147 and No.11874092) the Fok Ying-Tong Education Foundation, China (Grant No. 161005), the Planned Science and Technology Project of Hunan Province (Grant No. 2017RS3034), Hunan Provincial Natural Science Foundation of China (Grant No. 2016JJ2001), and Scientific Research Fund of Hunan Provincial Education Department (Grant No. 16B002).

**Reference**
[1] D. Jariwala, V. K. Sangwan, L. J. Lauhon, T. J. Marks, and M. C. Hersam, ACS Nano **8**, 1102 (2014).
[2] B. Anasori, M. R. Lukatskaya, and Y. Gogotsi, Nat. Rev. Mater. **2** (2017).
[3] D. Deng, K. S. Novoselov, Q. Fu, N. Zheng, Z. Tian, and X. Bao, Nat. Nanotechnol. **11**, 218 (2016).
[4] G. Xie, Z. Ju, K. Zhou, X. Wei, Z. Guo, Y. Cai, and G. Zhang, npj Computational Materials **4**, 21 (2018).
[5] K. S. Novoselov, A. Mishchenko, A. Carvalho, and A. H. Castro Neto, Science **353** (2016).
[6] W. Yu, Z. Zhu, S. Zhang, X. Cai, X. Wang, C.-Y. Niu, and W.-B. Zhang, Appl. Phys. Lett. **109**, 103104 (2016).
[7] K. Lai, C.-L. Yan, L.-Q. Gao, and W.-B. Zhang, J Phys. Chem. C **122**, 7656 (2018).
[8] K. Lai, H. Li, Y.-K. Xu, W.-B. Zhang, and J. Dai, Phys. Chem. Chem. Phys. (2019), 10.1039/C8CP07298A.
[9] Y. Cheng, Z. Zhu, W. Mi, Z. Guo, and U. Schwingenschlogl, Phys. Rev. B **87**, 100401 (2013).
[10] Y.-W. Son, M. L. Cohen, and S. G. Louie, Nature **444**, 347 (2006).
[11] W.-B. Zhang, Q. Qu, P. Zhu, and C.-H. Lam, J. Mater. Chem. C **3**, 12457 (2015).
[12] X. Li and J. Yang, J. Mater. Chem. C **2**, 7071 (2014).
[13] M. Ashton, D. Gluhovic, S. B. Sinnott, J. Guo, D. A. Stewart, and R. G. Hennig, Nano Lett. **17**, 5251 (2017).
[14] B. Huang, G. Clark, E. Navarro-Moratalla, D. R. Klein, R. Cheng, K. L. Seyler, D. Zhong, E. Schmidgall, M. A. McGuire, D. H. Cobden, W. Yao, D. Xiao, P. JarilloHerrero, and X. Xu, Nature **546**, 270 (2017).
[15] Z. Wang, I. Gutierrez-Lezama, N. Ubrig, M. Kroner, M. Gibertini, T. Taniguchi, K. Watanabe, A. Imamoglu, E. Giannini, and A. F. Morpurgo, Nat. Commun. **9**, 2516 (2018).
[16] D. R. Klein, D. MacNeill, J. L. Lado, D. Soriano, E. Navarro-Moratalla, K. Watanabe, T. Taniguchi, S. Manni, P. Canfield, J. Fern ández-Rossier, and P. Jarillo-Herrero, Science **360**, 1218

(2018).

[17] T. Song, X. Cai, M. W.-Y. Tu, X. Zhang, B. Huang, N. P. Wilson, K. L. Seyler, L. Zhu, T. Taniguchi, K. Watanabe, M. A. McGuire, D. H. Cobden, D. Xiao, W. Yao, and X. Xu, Science (2018), 10.1126/science.aar4851.

[18] H. H. Kim, B. Yang, T. Patel, F. Sfigakis, C. Li, S. Tian, H. Lei, and A. W. Tsen, Nano Letters **18**, 4885 (2018).

[19] S. S. Pershoguba, S. Banerjee, J. C. Lashley, J. Park, H. ˚Agren, G. Aeppli, and A. V. Balatsky, Phys. Rev. X **8**, 011010 (2018).

[20] K. L. Seyler, D. Zhong, D. R. Klein, S. Gao, X. Zhang, B. Huang, E. Navarro-Moratalla, L. Yang, D. H. Cobden, M. A. McGuire, *et al.*, Nat. Phys. **14**, 277 (2018).

[21] Y. Guo, S. Yuan, B. Wang, L. Shi, and J. Wang, J. Mater. Chem. C **6**, 5716 (2018).

[22] F. Zheng, J. Zhao, Z. Liu, M. Li, M. Zhou, S. Zhang, and P. Zhang, Nanoscale **10**, 14298 (2018).

[23] A. Avsar, J. Y. Tan, T. Taychatanapat, J. Balakrishnan, G. Koon, Y. Yeo, J. Lahiri, A. Carvalho, A. Rodin, E. OFarrell, *et al.*, Nature communications **5**, 4875 (2014).

[24] F. Katmis, V. Lauter, F. S. Nogueira, B. A. Assaf, M. E. Jamer, P. Wei, B. Satpati, J. W. Freeland, I. Eremin, D. Heiman, *et al.*, Nature **533**, 513 (2016).

[25] A. Molle, J. Goldberger, M. Houssa, Y. Xu, S.-C. Zhang, and D. Akinwande, Nature materials **16**, 163 (2017).

[26] R. Quhe, Y. Yuan, J. Zheng, Y. Wang, Z. Ni, J. Shi, D. Yu, J. Yang, and J. Lu, Sci. Rep. **4**, 5476 (2014).

[27] N. Gao, H. Liu, S. Zhou, Y. Bai, and J. Zhao, J Phys. Chem. C **121**, 5123 (2017).

[28] L. Y. B. H. Fengping Li, Wei Wei and Y. Dai, J Phys D: Appl Phys **50**, 115301 (2017).

[29] Y. Fan, X. Liu, J. Wang, H. Ai, and M. Zhao, Phys. Chem. Chem. Phys. **20**, 11369 (2018).

[30] L. Zhang, P. Bampoulis, A. N. Rudenko, Q. Yao, A. van Houselt, B. Poelsema, M. I. Katsnelson, and H. J. W. Zandvliet, Phys. Rev. Lett. **116**, 256804 (2016).

[31] Z. Ni, E. Minamitani, Y. Ando, and S. Watanabe, Phys. Chem. Chem. Phys. **17**, 19039 (2015).

[32] J. Zhu and U. Schwingenschlogl, J Mater. Chem. C **3**, 3946 (2015).

[33] M. Chen and M. Weinert, Phys. Rev. B **98**, 245421 (2018).

[34] M. Chen and M. Weinert, Nano letters **14**, 5189 (2014).

[35] G. Kresse and J. Hafner, Phys. Rev. B **47**, 558 (1993).

[36] G. Kresse and J. Furthmüller, Phys. Rev. B **54**, 11169 (1996).

[37] G. Kresse and D. Joubert, Phys. Rev. B **59**, 1758 (1999).

[38] J. P. Perdew, K. Burke, and M. Ernzerhof, Phys. Rev. Lett. **77**, 3865 (1996).

[39] S. Grimme, J Comput. Chem. **27**, 1787 (2006).

[40] H. J. Monkhorst and J. D. Pack, Phys. Rev. B **13**, 5188 (1976).

[41] W.-B. Zhang, Z.-B. Song, and L.-M. Dou, J. Mater. Chem. C **3**, 3087 (2015).

[42] G. Henkelman, A. Arnaldsson, and H. Jónsson, Computational Materials Science **36**, 354 (2006).

[43] M. X. Chen, Z. Zhong, and M. Weinert, Phys. Rev. B **94**, 075409 (2016).

[44] Y. R. Song, Y. Y. Zhang, F. Yang, K. F. Zhang, C. Liu, D. Qian, C. L. Gao, S. B. Zhang, and J.-F. Jia, Phys. Rev. B **90**, 180408 (2014).

[45] Y. S. Hou and R. Q. Wu, (2018), arXiv:1802.07358.

[46] H. Wang, F. Fan, S. Zhu, and H. Wu, EPL (Europhysics Letters) **114**, 47001 (2016).

[47] M. Bonilla, S. Kolekar, Y. Ma, H. C. Diaz, V. Kalappattil, R. Das, T. Eggers, H. R. Gutierrez, M.-H. Phan, and M. Batzill, Nat. Nanotechnol. **13**, 289 (2018).

Table 1 The calculated adsorption energy $E_b$ per M atom, in meV, and interlayer distance $d$, in Å, and charge transfer $\Delta\rho$, in e, from Mene to $CrX_3$, and exchange energy $\Delta E$, in meV, of $CrX_3$ in heterostructures, for each system at different stacking configurations.

| | Ge/CrI$_3$ | | | | Si/CrI$_3$ | | | | Si/CrBr$_3$ | | | |
|---|---|---|---|---|---|---|---|---|---|---|---|---|
| | $E_b$ | $d$ | $\Delta\rho$ | $\Delta E$ | $E_b$ | $d$ | $\Delta\rho$ | $\Delta E$ | $E_b$ | $d$ | $\Delta\rho$ | $\Delta E$ |
| Hol_up | 111 | 3.26 | 0.119 | 56 | 61 | 3.52 | 0.078 | 38 | 35 | 3.32 | 0.116 | 48 |
| Hol_dn | 102 | 3.27 | 0.108 | 58 | 64 | 3.63 | 0.067 | 40 | 38 | 3.40 | 0.141 | 51 |
| TbX_up | 150 | 3.04 | 0.115 | 61 | 95 | 3.17 | 0.082 | 43 | 55 | 3.08 | 0.125 | 52 |
| TbX_dn | 141 | 3.05 | 0.112 | 59 | 92 | 3.20 | 0.077 | 41 | 52 | 3.11 | 0.125 | 49 |
| TuX_up | 109 | 3.36 | 0.130 | 61 | 68 | 3.55 | 0.097 | 41 | 38 | 3.24 | 0.185 | 47 |
| TuX_dn | 115 | 3.23 | 0.142 | 53 | 64 | 3.50 | 0.091 | 36 | 43 | 3.27 | 0.123 | 53 |

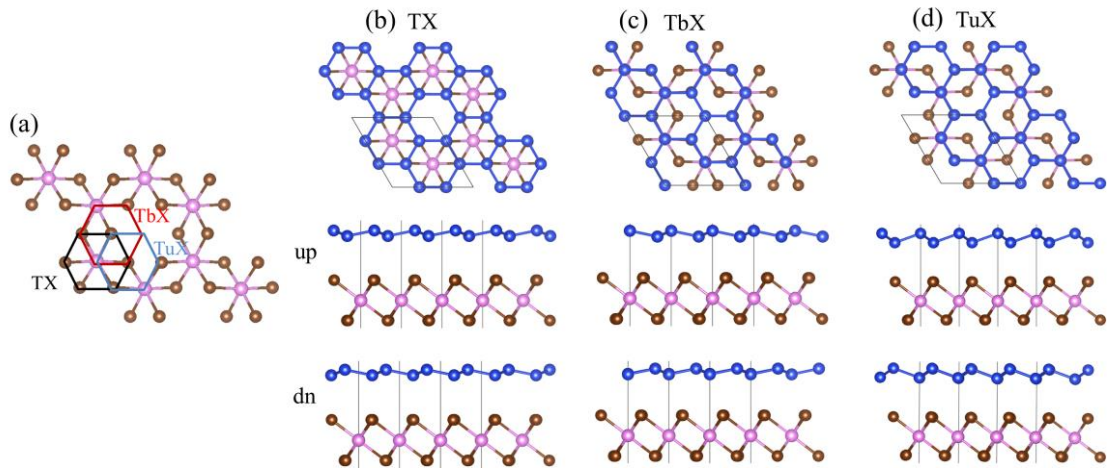

Figure 1 (a) Schematic diagram of three positions for hexagonal ring of Mene located on CrX3, (b)(c)(d) the first row, top views of heterostructures at position TX, TbX and TuX; the second and third row, side views of up and down M atom right above top X or Cr atom. The Cr, X, and M atoms are colored by pink, brown and blue.

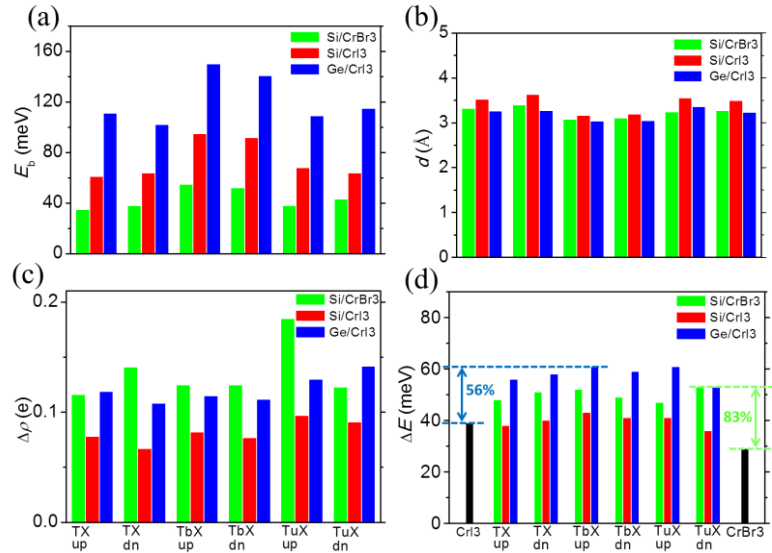

Figure 2 (a) The calculated adsorption energy $E_b$ per M atom, (b) interlayer distance $d$, (c) charge transfer from Mene to $CrX_3$ and (d) exchange energy $\Delta E$, for each system at different stacking configurations.

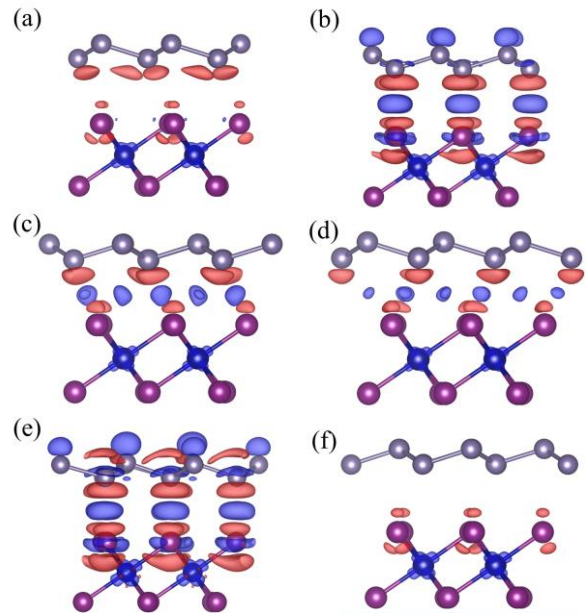

Figure 3 Charge difference density of $Ge/CrI_3$ for difference configuration (a)(b) TX_up/dn, (c)(d) TbX_up/dn, (e)(f) TuX_up/dn. The blue and red represent the charge accumulation and depletion, the isosurface is set as 0.0005 eV/Å$^3$.

Figure 4 The projected density of states (DOS) for Ge/CrI$_3$ at different configurations, (b)(f) TX_up/dn, (c)(g) TbX_up/dn, (d)(h) TuX_up/dn. (a)(e) the DOS of single-layer CrI$_3$ and germanene.

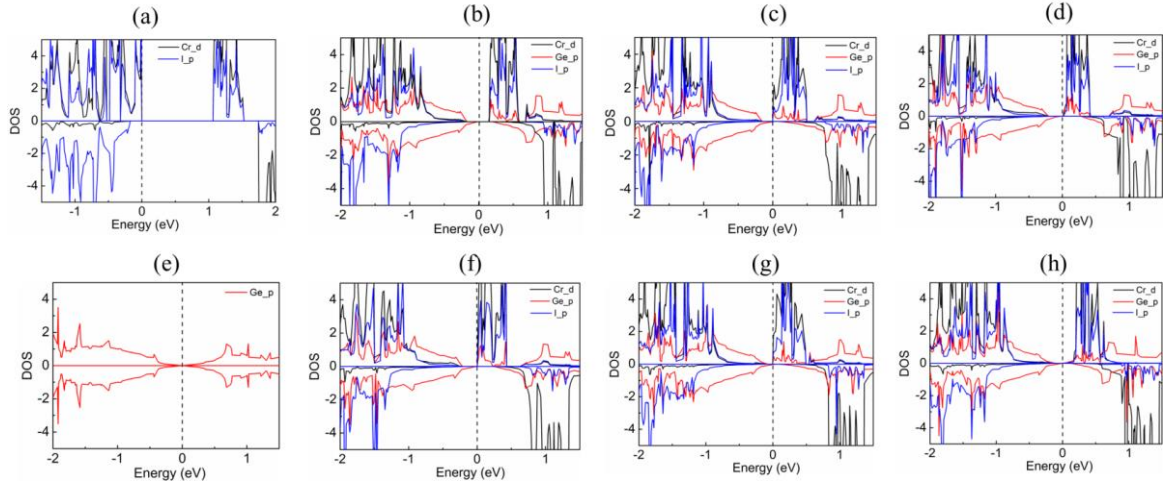

|  | TX_up | TX_dn | TbX_up | TbX_dn | TuX_up | TuX_dn |
|---|---|---|---|---|---|---|
| Si/CrBr$_3$ | 190 | 221 | 110 | 35 | 290 | 170 |
| Si/CrI$_3$ | 218 | 143 | 54 |  | 76 |  |
| Ge/CrI$_3$ | 165 | 25 | 200 | 76 | 215 |  |

Semiconductor — Half-metal — Metal

Figure 5 The characters of band structures for Ge/CrI$_3$, Si/CrI$_3$ and Si/CrBr$_3$ at different configurations. The number is the gap value, blue, green and red represent semiconductor, half-metal and metal, respectively